# Frozen Mode Regime in an Optical Waveguide With Distributed Bragg Reflector

Nathaniel Furman, Tarek Mealy, Md Shafiqul Islam, Ilya Vitebskiy, Ricky Gibson, Robert Bedford, Ozdal Boyraz, and Filippo Capolino

*Abstract*— We introduce a glide symmetric optical waveguide exhibiting a stationary inflection point (SIP) in the Bloch wavenumber dispersion relation. An SIP is a third order exceptional point of degeneracy (EPD) where three Bloch eigenmodes coalesce to form a so-called frozen mode with vanishing group velocity and diverging amplitude. We show that the incorporation of chirped distributed Bragg reflectors and distributed coupling between waveguides in the periodic structure facilitates the SIP formation and greatly enhances the characteristics of the frozen mode regime. We confirm the existence of an SIP in two ways: by observing the flatness of the dispersion diagram and also by using a coalescence parameter describing the separation of the three eigenvectors collapsing on each other. We find that in the absence of losses, both the quality factor and the group delay at the SIP grow with the cubic power of the cavity length. The frozen mode regime can be very attractive for light amplification and lasing, in optical delay lines, sensors, and modulators.

## I. INTRODUCTION

AN exceptional point of degeneracy (EPD) is a point where two or more of the eigenvalues and eigenvectors coalesce in the system parameter space of a given waveguide or structure [1]–[5]. The order of the EPD is given by the number of eigenmodes coalescing. For instance, the cutoff frequency in any uniform waveguide is a regular band edge (RBE), a second order EPD, as those in photonic crystals [6]. A degenerate band edge (DBE) is a fourth order EPD in a lossless and gainless waveguide [6]–[9]. In the past decade, most of the literature was focused on EPDs obtained contingent upon parity-time (PT) symmetric systems, i.e., with balanced gain and loss [10]–[20]. However, the occurrence of EPDs does not require PT symmetry [6], [8], [20], [21].

This research focuses on a third order EPD, called the stationary inflection point (SIP), that occurs to optical modes in waveguides in the absence of gain and loss [22]. In particular, we focus on a periodic, glide symmetric [23], [24], reciprocal, lossless photonic optical waveguide (OWG) with an SIP, where three modes coalesce to form the so called "frozen-mode". Here, the optical waveguide is made of three paths. Other waveguides showing the SIP in periodic lossless and gainless gratings and photonic crystals are in [8], [25]–[29].

Other research has shown two-path non-reciprocal structures supporting the SIP using magnetic materials to break system reciprocity [30]–[33].

A microwave study and experimental confirmation of a similar three-path waveguide supporting the SIP was presented in [29], and a study in the optical regime supporting the SIP was presented in [8] and [28], [34].

In this paper, a novel geometry is presented that supports the SIP, which is verified in two ways: through observation of the dispersion relationship and via calculation of the coalescence parameter. This research (with the unit cell in Fig. 1) differs from previous similar work like the CROW [8] and ASOW [34] in a few key aspects. This three-path OWG incorporates a distributed Bragg reflection (DBR), used to interfere, or mix, $+z$ traveling and $-z$ traveling modes necessary to couple three waves that form the SIP. We also observe that the ports can be easily controlled with many options for input, output, and termination conditions. Furthermore, the radius of the bent Silicon waveguide used in this paper is approximately $2\,\mu m$, much smaller than those in [8], [34] using different geometries, chosen for a small device footprint. Most importantly, the smaller unit cell used in this paper is useful to further separate the frequencies of the SIP and of RBEs (see Fig. 2a), compared to the frequency separation achieved using the geometries in [8], [34]. The OWG in this paper can support the SIP with both larger and smaller radii. Any of the six ports of the finite length structure can be taken as the input or as the output of the cavity, and each port can be terminated with a unique load or boundary condition. These features offer high flexibility based on fabrication restrictions, power extraction, impedance matching, and more.

A key property of the SIP is the significant reduction in the group velocity of waves propagating through the waveguide. This EPD, again obtained through the coalescence of three eigenmodes, forms a frozen mode at the SIP frequency where the group velocity vanishes [1], [7], [35]. At frequencies slightly adjoint to the SIP frequency, the direction of propagating modes remains consistent, i.e., the propagating modes do not show an inversion of the group velocity as instead happens near the RBE and DBE that separate a pass band from a stop band. The SIP's group velocity direction consistency has applications in optical switching, lasers [36], delay lines [28], [37], and high-power amplifiers [38] where the SIP is used to facilitate strong interaction with electron beam charges by slowing down the electromagnetic waves and obtaining high output gain.

In Sec. II, we discuss the unit cell made of two coupling segments and two chirped DBR (cDBR) segments. In this section, we also discuss the sub-segment blocks used to model the entire structure. In Sec. III, we discuss optimization methods for unit cell dimensions and parameters based on the three eigenmode coalescence. In Sec. IV, we show various representations of the dispersion diagram of the modes in the infinitely long periodic waveguide supporting the SIP. In Sec. V, we show the transfer function for a finite length waveguide and discuss wave properties.

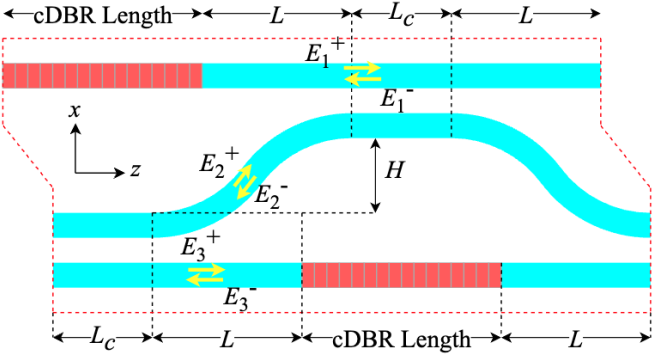

Fig. 1: The unit cell of the three-path waveguide which supports the SIP. The structure is periodic with period $d = 2L + L_c + L_{cDBR}$. The effective field coupling coefficient between the straight waveguides is $|\kappa|$ and the cDBR has a reflection coefficient magnitude $\rho_0$ at the SIP frequency. The figure shows the orientation of the electric field wave amplitudes along the $+z$ and $-z$ propagating directions.

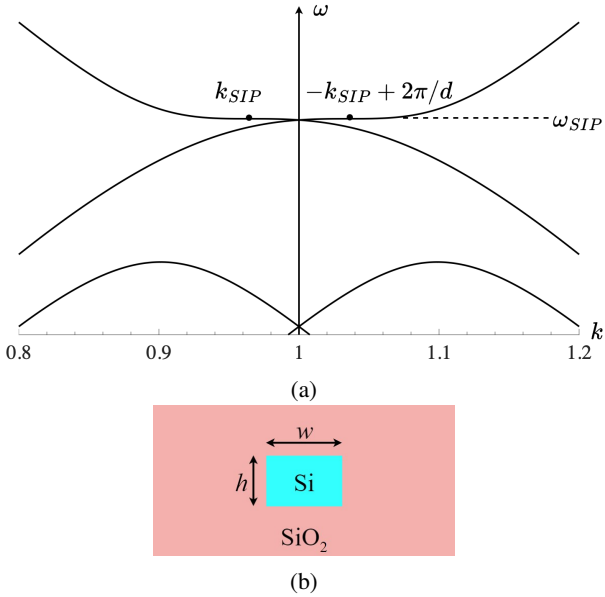

Fig. 2: (a) Dispersion of the purely-real Bloch wavenumber of the propagating modes in the optical waveguide with unit cell as in Fig. 1, showing the SIP at angular frequency $\omega_{SIP} = 2\pi f_{SIP}$. The modes with complex wavenumber are shown later in the paper. (b) General representation of the silicon core waveguide with silicon oxide cladding with height $h = 220\,\text{nm}$ and width $w = 450\,\text{nm}$.

## II. FIELD DESCRIPTIONS AND GEOMETRY

The periodic three-way waveguide proposed in this paper has the unit cell in Fig. 1 that consists of two chirped DBRs (one for each the top and bottom OWG) and two directional couplers (one between each of the two upper and two lower OWGs) to realize the SIP.

Four parameters are used to describe the waveguide unit cell dimensions. As will be discussed more in Appendix C, the cDBR (with length $L_{cDBR}$) is fixed for a given design. The three parameters which we vary when looking for an SIP are the length of the coupled sections, the length of the straight sections connecting the cDBR to the coupling sections, and the length between the top and bottom parts of the middle waveguide, denoted by $L_c$, $L$, and $H$, respectively in Fig. 1.

We use an Si strip in $SiO_2$ cladding architecture as in Fig. 2b, with refractive indexes $n_{Si} = 3.48$ and $n_{SiO_2} = 1.45$, respectively, at 193 THz. The waveguide cross section has the dimensions $h = 220\,\text{nm}$ by $w = 450\,\text{nm}$. The dimensions chosen are common in multiple foundry standard processes development kits. For simplicity, we neglect material dissipation and sidewall scattering losses.

The time convention $e^{j\omega t}$ is implicitly assumed throughout this paper. We represent the electromagnetic wave propagating along the $z$-direction using the complex electric field amplitudes (phasors) moving in the positive-$z$ and negative-$z$ propagating directions as $\mathbf{E}_n(z) = \left[ E_n^+(z), \; E_n^-(z) \right]^T$ where the superscript $T$ denotes the transpose operation. Thus, the wave amplitudes in the three waveguides in Fig. 1 are represented using the six-dimensional state vector

$$\boldsymbol{\Psi}(z) = \begin{pmatrix} \mathbf{E}_1(z) \\ \mathbf{E}_2(z) \\ \mathbf{E}_3(z) \end{pmatrix} \tag{1}$$

similar to the formalism in [6]–[8], [39], [40].

The analysis procedure performed here is similar to the one in [8] based on transfer matrices. In this paper, the transfer matrix is either directly derived using analytic expressions or calculated from the scattering matrix evaluated via full-wave simulations. The transfer matrix relates the state vector at location $z_1$ to $z_2$ as $\boldsymbol{\Psi}(z_2) = \underline{\mathbf{T}}(z_2, z_1)\boldsymbol{\Psi}(z_1)$. Specifically for the evolution of the state vector across the unit cell of length $d$, we have

$$\boldsymbol{\Psi}(z + d) = \underline{\mathbf{T}}_U \boldsymbol{\Psi}(z) \tag{2}$$

where $\underline{\mathbf{T}}_U$ is the transfer matrix of the unit cell. From Floquet-Bloch theory, we look for periodic solutions of the state vector as $e^{-jkd}$ where $k$ is the Floquet-Bloch complex wavenumber. These solutions must satisfy $\boldsymbol{\Psi}(z + d) = \zeta \boldsymbol{\Psi}(z)$ where $\zeta \equiv e^{-jkd}$, leading to the eigenvalue problem

$$\underline{\mathbf{T}}_U \boldsymbol{\Psi} = \zeta \boldsymbol{\Psi}. \tag{3}$$

The six eigenvalues $\zeta_n \equiv e^{-jk_n d}$, with $n = 1, 2, ...6$, are obtained by solving the dispersion characteristic equation $D(k, \omega) \equiv \det\left[\underline{\mathbf{T}}_U - \zeta \underline{\mathbf{1}}\right] = 0$ with $\underline{\mathbf{1}}$ representing the $6 \times 6$ identity matrix. Because this waveguide is reciprocal, the determinant of the transfer matrix satisfies $\det\left[\underline{\mathbf{T}}_U\right] = 1$. This means the dispersion diagram shows a symmetry such that if $k(\omega)$ is a solution of Eq. (3), then $-k(\omega)$ is as well. Hence, the dispersion diagram is symmetric with respect to the boundary of the Brillouin Zone (BZ), defined here with $\mathrm{Re}(k)$ from $-\pi/d$ to $+\pi/d$.

At an SIP, three wavenumbers are equal to each other, i.e., $k_1 = k_2 = k_3 = k_{SIP}$. Therefore, because of reciprocity, we also have an SIP at $k_4 = k_5 = k_6 = -k_{SIP}$, which, because of periodicity, induces an SIP also at $-k_{SIP} + 2\pi/d$, as shown in Fig. 2a.

### A. Phase Delay Subblocks

This section describes the phase delay of the electric fields for the straight waveguide of length $L$ and the curved portion of the waveguide with arcs of radius $R_2$ (see Fig. 3). For a given length $z$, the propagation delay for a $+z$ traveling wave is $\Omega_{w+} = e^{-jk_w z}$. Here, $k_w$ is the propagation wavenumber in the straight waveguide expressed as $k_w = n_w k_0$. The modal refractive index of the quasi-TE mode in the straight rectangular waveguide is $n_w$ and $k_0$ is the free space wavenumber. Based on the waveguide dimensions, $n_w = 2.351$ at 193 THz. This value was calculated using the port analysis of a frequency domain electromagnetic simulation software called CST Studio Suite based on the finite element method (FEM). For a wave traveling in the opposite, or $-z$, direction the sign in the exponential is positive and denoted as $\Omega_{w-} = e^{+jk_w z}$. Using this notation, we write the phase delay transfer matrix as

$$\underline{\mathbf{T}}_{phase} = \begin{pmatrix} \Omega_{w+} & 0 \\ 0 & \Omega_{w-} \end{pmatrix}. \qquad (4)$$

This transfer matrix subblock is used to describe the phase propagation in the straight waveguide segments with length $L$. We also use the same transfer matrix with the different propagation distance $2\alpha R_2$ for describing wave propagation in the two curved section of the middle waveguide in Fig. 3. We build the path in the middle waveguide out of four identical arcs translated around the unit cell. To build these arcs, we write equations describing the intersection of two circles centered at point $O$ and point $P$ in Fig. 3. If we set the intersection point at $(L'/2, H/2)$ where $L' = L + L_{cDBR}/2 - L_c/2$, we calculate the required radius of the second circle as

$$R_2 = \frac{(H+w)^2 + L'^2}{4(H+w)}. \qquad (5)$$

Using this formulation, the curved section of the middle waveguide makes smooth transitions between its lower and upper paths with the maximum allowable radius. This helps reduce mode mismatch losses and losses associated with decreasing radii of curvature. Using $R_1 = \sqrt{(L'/2)^2 + [(H+w)/2]^2}$ and letting $\alpha$ represent the arc angle, we calculate $\alpha = 2\arcsin[R_1/(2R_2)]$.

### B. cDBR Subblock

When considering the cDBR of a given length $L_{cDBR}$, we impose three requirements on the scattering matrix describing the field amplitudes for an idealized model. These requirements are for a reciprocal, symmetric, and lossless system.

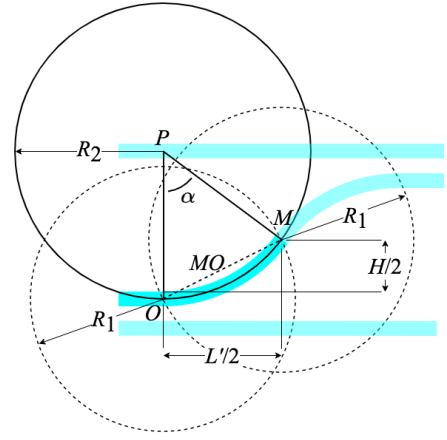

Fig. 3: Geometric description of the bottom left middle curved waveguide arc of radius $R_2 = 2.12\ \mu\mathrm{m}$ and length 3.01 $\mu\mathrm{m}$.

Although full-wave models incorporate loss (discussed more in Appendix C), they can be well-approximated as lossless because of the minimal losses associated with the cDBR. Full-wave models are almost perfectly reciprocal and symmetric, with any deviations from a perfectly reciprocal and symmetric model resulting from finite numerical precision. Thus, we examine the properties of a reciprocal, symmetric, and lossless scattering matrix used to model the cDBR subblock.

As classically known, $S_{11} = S_{22}$ and $S_{12} = S_{21}$ for a two port symmetric and reciprocal system. The unitary condition for a lossless network is summarized by the two equations: $1 = S_{11}S_{11}^* + S_{21}S_{21}^*$ and $0 = S_{11}S_{21}^* + S_{21}S_{11}^*$. Letting $S_{11} = \rho_0 e^{j\theta_\rho}$ and $S_{21} = \tau_0 e^{j\theta_\tau}$, the first equation leads to $\tau_0 = \sqrt{1 - \rho_0^2}$, whereas the second equation results in $\theta_\tau = \theta_\rho \pm \pi/2$. Therefore, for a lossless, or unitary, scattering matrix, $S_{21} = \sqrt{1 - \rho_0^2}\, e^{j(\theta_\rho - \pi/2)}$. We have chosen to subtract $\pi/2$ from $\theta_\rho$ for consistency with the full-wave simulation results.

We choose a frequency dependent magnitude and phase for the reflection coefficient $S_{11}$ to approximate the cDBR in a lossless analytical representation. The chosen reflection coefficient magnitude and phase are based on the FEM implemented by CST Studio Suite full-wave frequency domain simulations described further in Appendix C. Thus, in the following we use *two models* for the cDBR to build the transfer matrix of a unit cell: (i) the analytic model and (ii) the full-wave cDBR model. (All the other sub blocks are based on an analytic description of transfer matrix subblocks). The *analytic model* is completely lossless and uses the transfer matrix given as

$$\underline{\mathbf{T}}_\rho = \frac{1}{\tau_0 e^{j\theta_\tau}} \begin{pmatrix} (\tau_0 e^{j\theta_\tau})^2 - (\rho_0 e^{j\theta_\rho})^2 & \rho_0 e^{j\theta_\rho} \\ \rho_0 e^{j\theta_\rho} & 1 \end{pmatrix}. \qquad (6)$$

This $T$ matrix is obtained by applying the relationships described in Appendix B, to convert the symmetric, reciprocal, and lossless simulated scattering matrix to the transfer matrix, so that it describes the ideal lossless cDBR subblock.

The *full-wave cDBR model* uses the simulated scattering parameters from the frequency domain solver transformed to the transfer matrix using equations in Appendix B. This scattering matrix from simulations does not fully satisfy the unitary conditions and thus incorporates the small loss from the cDBR into the full-wave model.

### C. Distributed Coupler Subblock

Literature commonly documents the well-known beam splitter relationships, particularly for a 3-dB coupler [41], [42]. However, the matrices and equations describing the coupling between two waveguides are generally given as point-like coupling. For some geometries, such as the CROW, this assumption is reasonable for small field coupling coefficients as the gap between waveguides can be adjusted accordingly. However, we must explicitly take into account the distributed nature of our coupling. We perform the analysis in Appendix D using even and odd mode electric field profiles. In our geometry, we set the gap size to $150\,\mathrm{nm}$. Once we determine the even and odd mode effective refractive indices as $n_e$ and $n_o$ through full-wave simulations, we write the full distributed coupling transfer matrix as

$$\underline{\mathbf{T}}_{couple} = \begin{pmatrix} \frac{\Omega_e + \Omega_o}{2} & 0 & \frac{\Omega_e - \Omega_o}{2} & 0 \\ 0 & \frac{\Omega_e + \Omega_o}{2\Omega_e \Omega_o} & 0 & \frac{\Omega_e - \Omega_o}{2\Omega_e \Omega_o} \\ \frac{\Omega_e - \Omega_o}{2} & 0 & \frac{\Omega_e + \Omega_o}{2} & 0 \\ 0 & \frac{\Omega_e - \Omega_o}{2\Omega_e \Omega_o} & 0 & \frac{\Omega_e + \Omega_o}{2\Omega_e \Omega_o} \end{pmatrix} \quad (7)$$

where $\Omega_e$, $\Omega_o$ take the same form as in Eq. (4) for forward (positive $z$) phase propagation over a length $L_c$.

### D. Combining the Subblocks of a Unit Cell

We combine the proper subblock transfer matrices in a full $6 \times 6$ transfer matrix describing the three waveguides for each section. The unit cell is broken into four sections, two describing the purely phase propagation elements ($L$ and the curved parts of the middle OWG) and two describing the coupling and cDBR as seen in Fig. 1. The sections comprised of the length $L$ also include the arcing portions of the middle waveguide. Thus, these two sections are comprised of purely phase delay elements. The $T$ matrix $\underline{\mathbf{T}}_U$ of the entire unit cell is then calculated by multiplying the $T$ matrices of the different sections in the proper order.

We reiterate here that this work compares two models of the waveguide as described in subsection II-B: the "analytic model" and the "full-wave model". The only difference between the two models is regarding the cDBR modeling. Both models rely on full-wave FEM frequency domain simulations of the chirped distributed Bragg reflector. The analytic model extracts from the full-wave model of the cDBR the average of the reflection coefficient magnitude and phase over the frequency range of interest and calculates the transmission coefficient magnitude and phase such that the scattering matrix satisfies the lossless condition. The resulting transfer matrix is given in Eq. (6). The full-wave model instead computes the $T$ matrix directly from the lossy $S$ matrix obtained by the full-wave simulations using the procedure in Appendix B, i.e., it does not modify the scattering parameters from FEM simulations as instead done by the analytical model. As a result, the analytic model is lossless whereas the full-wave model incorporates small losses associated with the cDBR.

### III. DESIGN OPTIMIZATION USING THE COALESCENCE PARAMETER

After creating the unit cell by combining the various sub-blocks in the waveguide, we turn to choosing the geometric parameters which support the SIP. We discuss here the optimization of the unit cell that leads to an SIP and examine how changes to the geometric values minimize the coalescence parameter. The coalescence parameter $C$ is a figure of merit for evaluating EPD conditions and measures the sum of the angles between the eigenwavevectors of Eq. (1) in a six-dimensional space. It is a measurement of how close a system is to an EPD of a given order. As presented in [29], the coalescence parameter for an SIP is calculated by

$$C = \frac{1}{3} \sum_{\substack{m=1, n=2 \\ n>m}}^{3} |\sin\theta_{mn}|, \quad \cos\theta_{mn} = \frac{|\langle \mathbf{\Psi}_m, \mathbf{\Psi}_n \rangle|}{||\mathbf{\Psi}_m|| \, ||\mathbf{\Psi}_n||} \quad (8)$$

where $\theta_{mn}$ is the angle between the two six-dimensional normalized complex state vectors. In this equation, $\mathbf{\Psi}_1$, $\mathbf{\Psi}_2$, and $\mathbf{\Psi}_3$ represent the three eigenvectors associated with the three wavenumbers such that $0 < k_n < \pi/d$ in Eq. (3). The inner product is defined as

$$\langle \mathbf{\Psi}_m, \mathbf{\Psi}_n \rangle = \mathbf{\Psi}_m^\dagger \, \mathbf{\Psi}_n. \quad (9)$$

The dagger symbol, $\dagger$, represents the complex conjugate transpose operation and $||\mathbf{\Psi}_m||$, $||\mathbf{\Psi}_n||$ denote the vectors' norms. The coalescence parameter is always positive and less than one, and $C = 0$ indicates perfect coalescence of the three eigenvectors, i.e., the system experiences the SIP, which is an EPD of order three. Thus, the geometric parameters are chosen to minimize $C$.

The waveguide exhibits the SIP with $L = 1.738\,\mu\mathrm{m}$, $L_c = 2.075\,\mu\mathrm{m}$, and $H = 3.152\,\mu\mathrm{m}$. These parameters result in a bend radius of $2.12\,\mu\mathrm{m}$ per Eq. (5). We use the cDBR with a length of $L_{cDBR} = 7\,\mu\mathrm{m}$ as detailed further in Appendix C.

### IV. DISPERSION DIAGRAM OF EIGENMODES

In Fig. 4 we show the complex-wavenumber dispersion diagram of the eigenmodes in the waveguide. The effective field coupling coefficient based on a distributed length $L_c$ is $|\kappa| \approx 0.214$ and the magnitude of the field reflection coefficient of the simulated cDBR is $\rho_0 \approx 0.517$ at $193\,\mathrm{THz}$. In Fig. 4a we show the complex dispersion diagram of the analytic model with purely real $k$ (or propagating) modes in black and complex $k$ (or evanescent) modes in red. For $\mathrm{Re}(kd/\pi) > 1$, we see a positive slope for propagating modes whereas there is a negative slope for $\mathrm{Re}(kd/\pi) < 1$. Therefore, around $k_{SIP}$, the propagating mode is backward, i.e., the phase and group velocities have opposite directions. Over the frequency range of interest, all the propagating-mode curves (black) are monotonically increasing or decreasing, a key property of the SIP. Additionally, at the SIP, we see that all modes are purely real with $\mathrm{Im}(kd/\pi) = 0$. Fig. 4b compares the analytic model using a lossless representation of the chirped DBR to the full-wave cDBR model which uses the scattering parameters from simulations directly. The small losses associated with the cDBR account for the slight differences in the two models.

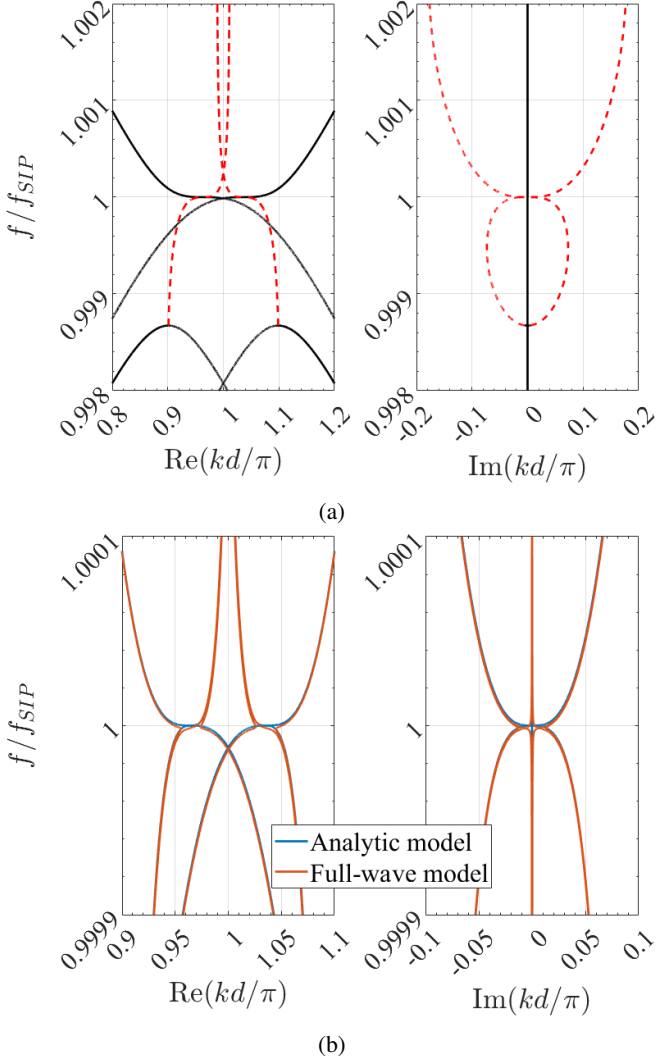

(a)

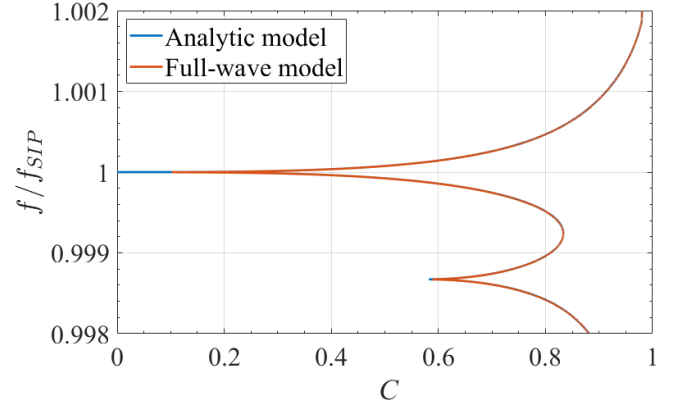

Fig. 5: Coalescence parameter calculated for the analytic and full-wave model representing the degree to which three eigenvectors are parallel with respect to each other. The near-zero minimum indicates an SIP where the local minimum (at $f/f_{SIP} \approx 0.9986$) indicates a lower-order EPD; in this case an RBE.

(b)

Fig. 4: (a) The complex dispersion diagram for the analytic model, with propagating modes in black and evanescent modes in red, where $f_{SIP} = 193$ THz. (b) Zoomed in complex dispersion relationship for the analytic model in blue and the cDBR model in orange. The minimal differences in the models are due to the very small losses associated with the cDBR.

In Fig. 5 we show the coalescence parameter based on the eigenvectors calculated using the two models. The analytic model shows a near-perfect $C = 0$ coalescence (i.e., the three eigenvectors are almost perfectly coalesced). Because of small radiation losses associated with the cDBR, the coalescence parameter for the full-wave models cannot converge to zero. That means that the three eigenvectors are not fully coalesced, hence the SIP is slightly perturbed. However, we still see very close agreement between the two models, and still, the case including losses in the DBR shows a good degree of eigenvector coalescence.

Two additional useful representations of the wavenumber dispersion diagram are in Fig. 6. In Fig. 6a, we plot the normalized wavenumber in complex space obtained from the analytic model where the arrows represent increasing frequency and the colored lines are the same as in Fig. 4a.

The red dashed lines represent evanescent modes, whereas the solid black lines represent the propagating (i.e., with purely real $k$) modes. The three modes with $\mathrm{Re}(k) < \pi/d$ share the same wavenumber $k_{SIP}$ as well as the three modes with $\mathrm{Re}(k) > \pi/d$ share the same wavenumber $-k_{SIP} + 2\pi/d$ at the SIP frequency. Around each SIP, the curves are offset $120°$ with respect to each other directly before and after the SIP frequency. Fig. 6b shows $\log_{10} |D(k, \omega)|$ for real wavenumbers (i.e., we impose $\mathrm{Im}(k) = 0$). The two propagating branches are shown in the yellow-blue overlay plane; they correspond to the black curves in Fig. 4a. In the three dimensional plot we can clearly see the two SIPs at $k_{SIP}$ and $-k_{SIP} + 2\pi/d$, represented by the two dips at $f = f_{SIP}$. This figure is particularly interesting because, while all blue points on the overlay plane result in a zero determinant (i.e. the points solve the dispersion characteristic equation $D(k, \omega) = 0$), the frequency-wavenumber pair at the SIP more closely solves the dispersion equation. In other words, the triple eigenmode degeneracy at the SIP results in the dispersion characteristic equation approaching zero much more rapidly than when not at the SIP. In [34], it is shown that in the proximity of $k_{SIP}$ the dispersion equation is $D(k, \omega_{SIP}) \propto (k - k_{SIP})^3$ showing that the way $D$ approaches zero is much faster near an SIP. Analogous behavior happens near $-k_{SIP} + 2\pi/d$. This observation explains why the plot in Fig. 6b has the two very pronounced dips.

In the vicinity of the SIP at $k_{SIP}$, the dispersion is well approximated by

$$\omega - \omega_{SIP} \approx \eta(k - k_{SIP})^3 \qquad (10)$$

where $\eta$ is a constant describing the flatness of the real $k$ branch (i.e., the propagating mode) at the SIP, analogous to the theory presented in [9]. This constant $\eta$ is related to the third derivative of $\omega$ with respect to $k$ at the SIP angular frequency $\omega_{SIP}$. The group velocity and its derivative are zero at the SIP, written as $\partial\omega/\partial k = 0$ and $\partial^2\omega/\partial k^2 = 0$ whereas the second derivative of the group velocity is nonzero, which is $\partial^3 \omega / \partial k^3 = 6\eta \neq 0$.

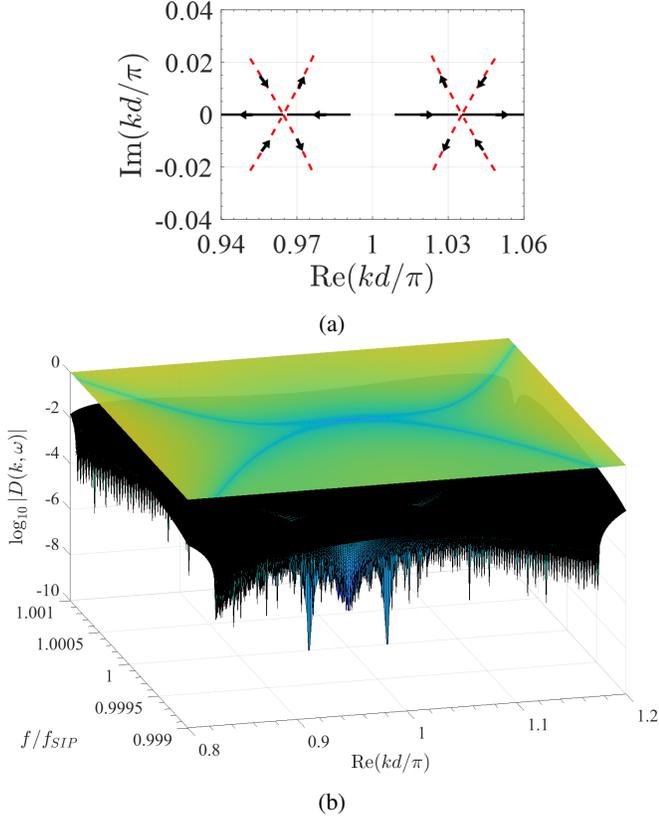

(a)

(b)

Fig. 6: (a) Imaginary versus real wavenumbers of the dispersion diagram for the analytic model with arrows representing the direction of increasing frequency, red lines representing evanescent modes, and black lines representing propagating modes. (b) Contour of the dispersion characteristic equation. The blue curve in the yellow-blue overlay represent solutions to the characteristic equation, e.g. a determinant of zero. At the SIP, there are triple poles which result in the determinant approaching zero much quicker than not at the SIP, as is clearly shown in the figure. Due to finite step sizes in frequency and wavenumber, the 3D contour of the blue curve is not perfectly zero.

We briefly mention three interesting aspects of the dispersion diagram: the deviation of the SIP wavenumber from the BZ boundary, less than $0.05\pi/d$ (Fig. 4a on the left and Fig. 2a); the Dirac-like intersection of spectral real $k$ branches at the BZ boundary, at a frequency slightly lower than the SIP frequency; and the next closest RBE at a frequency lower than the SIP frequency, shown in Fig. 4a and Fig 5. The small deviation of the real parts of the SIP wavenumber from the BZ boundary may be especially beneficial to sensing applications. Despite the vicinity of the two 3rd order EPDs, the intersection is not related to a sixth order degeneracy, or 6DBE [43]. At frequencies slightly less than the SIP frequency, the dispersion diagram shows two spectral real $k$ branches intersecting each other (see Fig. 2a). This spectral branch intersection is typical of modes in waveguides with glide symmetry [23], [24]. The intersection so close to the SIP may affect the frozen mode regime and cause behaviors that have yet to be explored.

As seen in Fig. 4a and Fig. 5, the next closest EPD to the SIP is an RBE at $f_{RBE}/f_{SIP} \approx 0.9986$, or more than $250\,\mathrm{GHz}$ away from the SIP in non-relative frequency units. This separation between EPDs is significant and may help isolate the SIP behavior from other EPD effects. More work is needed to completely understand the effects each of these observations has on a finite length structure operating in the frozen mode regime.

## V. SIP Cavity Analysis

We analyze how the transfer function, reflection function, group delay, and quality factor of the cavity vary with an increasing number of cascaded unit cells. At the extreme left and right sides of the given number of unit cells $N$, we include an additional cell, on each side, described by Fig. 7a and Fig. 7b. We include these additional geometries as to not start or terminate the structure with the cDBR and coupling blocks. The left edge of Fig. 7a and the right edge of Fig. 7b do not include the cDBR or additional coupling segments. The transfer matrix of the leftmost cell is denoted as $\underline{\mathbf{T}}_{aux, A}$, the rightmost as $\underline{\mathbf{T}}_{aux, B}$, and both are calculated using analytic methods as described in Sec. II. In the finite-length structure, we take the input and output from the top waveguide and thus define the transfer function as

$$T_f = \frac{E_1^+(Nd + d')}{E_1^+(0)} \tag{11}$$

where $N$ is the number of unit cells of length $d$ as shown in Fig. 1, and $d' = 4L + 2L_{cDBR} + L_c$ is the added length from the two auxiliary geometries. We assume the output of the top waveguide is terminated by the characteristic impedance of the unperturbed waveguide so as to not induce additional reflections from outside the structure, i.e. matched loads to the single-waveguide impedance (but not matched to the Bloch impedance). We then define the reflection function as

$$R_f = \frac{E_1^-(0)}{E_1^+(0)}. \tag{12}$$

We assume that the boundary conditions applied to the four terminations (two on the left and two on the right) of the middle and bottom waveguides are ideal reflections from open ended waveguides, i.e., with a reflection coefficient of $\Gamma = +1$.

To calculate the transfer function and reflection function, we cascade the $T$ matrix of a single unit cell $N$ times and find the state vector

$$\boldsymbol{\Psi}(z = Nd + d') = \underline{\mathbf{T}}_{aux, B}\,\underline{\mathbf{T}}_U^N\,\underline{\mathbf{T}}_{aux, A}\,\boldsymbol{\Psi}(z = 0). \tag{13}$$

Here, $\underline{\mathbf{T}}_U$ is the transfer matrix for a single unit cell using the analytic model. Fig. 8 and Fig. 9 show the transfer function and reflection magnitudes for waveguides made by five different numbers of unit cells. The given $N$ represents the number of unit cells, not including the two auxiliary geometries. As the number of unit cells increase, we see the transfer function magnitude approaches unity at the SIP frequency. Similarly,

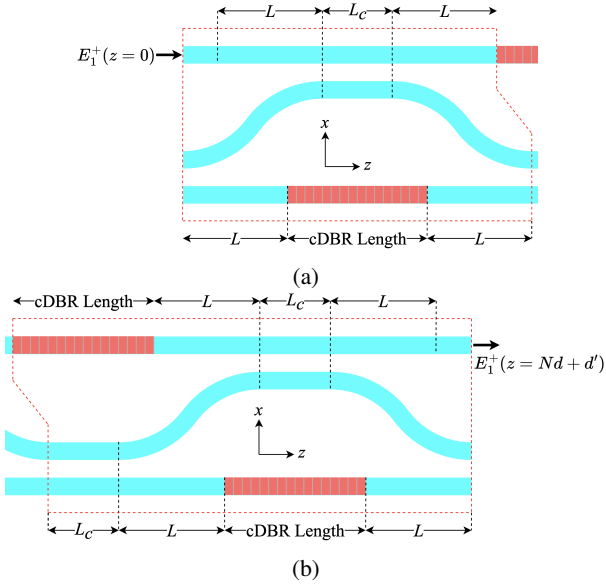

Fig. 7: Auxiliary waveguide cells for (a) the start of the finite length cavity and (b) the end of the finite length cavity. They look similar to the periodic unit cells but they are slightly different. In (a), the bottom two waveguides are uncoupled to each other and the top waveguide is without the cDBR on the left edge. In (b), the top waveguide does not include the cDBR and the bottom two waveguides are not coupled to each other on the right side.

the reflection function magnitude follows the trend the transfer function establishes.

An important application of the SIP is in optical delay lines [28]. In that application, the group delay and quality factor are useful parameters in characterizing the response of the waveguide. We use the definition of group delay as the negative derivative of the transfer function phase with respect to the angular frequency as given by

$$\tau_g = -\frac{\partial \angle T_f(\omega)}{\partial \omega}. \tag{14}$$

Fig. 10 shows that with an increasing number of unit cells, the group delay at the SIP resonance frequency scales proportionally to $N^3$ and can be described by $aN^3 + b$ where $a = 5.1$ fs and $b = 11.1$ ps for this geometry. For an $N = 30$ unit cell structure, the group delay at the SIP is 156 ps. In an isolated *straight* waveguide of equivalent length, the propagation delay is only $\tau_s = 3.1$ ps. This delay is calculated by $\tau_s = (Nd+d')/v_{ph}$ where $v_{ph} \approx 1.27 \times 10^8$ m/s is calculated using the same modal index in Sec. II-A and $N$, $d$, and $d'$ are as defined before in Eq. (11). The group delay at the SIP frequency is nearly 50 times greater than that in a straight, isolated, waveguide of the same length. Alternatively, we can describe the delay by an effective group refractive index $n_{eff,g}$ calculated from the group delay and the total waveguide length. For $N = 30$, we have $n_{eff,g} = c_0 \tau_g /(d+d') = 117$.

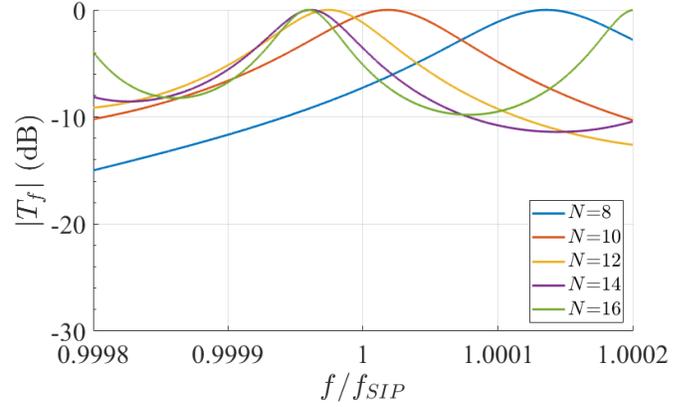

Fig. 8: Magnitude of the transfer function in dB of the waveguide near the SIP frequency calculated using analytic methods for waveguides with five different numbers of unit cells $N$. This $N$ does not include the two auxiliary geometries at the left and right ends.

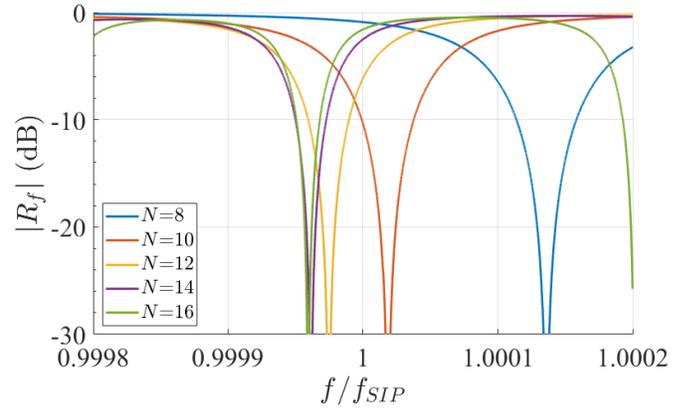

Fig. 9: Magnitude of the reflection function in dB of the waveguide near the SIP frequency calculated using analytic methods at five different numbers of unit cells $N$.

The quality factor of the waveguide is well approximated by

$$Q = \frac{\omega_{SIP}\tau_g}{2} \tag{15}$$

where $\omega_{SIP}$ is the angular SIP frequency and is shown in Fig. 11. Similar to the group delay, we see a third order scaling law with the number of unit cells.

## VI. CONCLUSION

We have detailed the design of a glide-symmetric three-path optical waveguide supporting the frozen mode regime associated with the stationary inflection point (SIP) of the Bloch dispersion relation. The proposed design involves a directional coupler and a chirped distributed Bragg reflector. The SIP is realized at 193 THz, or at the free space wavelength of $\approx 1550$ nm. The concept of coalescence parameter has been used to optimize the unit cell geometric parameters and verify the SIP behavior in both the analytic model and the full-wave model using simulations calculated using the finite element method in a frequency domain electromagnetic wave solver. At the SIP frequency, the transfer function of a finite structure is close to unity as the number of unit cells is increased. Both the group delay and the quality factor of the finite structure grow as the third power of the number of unit cells. The frozen mode regime can be used in low-dispersion optical delay lines, for light amplification, and for lasing. In addition, the proximity of the pair of reciprocal SIPs to each other and to the degeneracy point at the Brillouin zone boundary seen in Fig. 4 can be particularly attractive for light modulation and sensing.

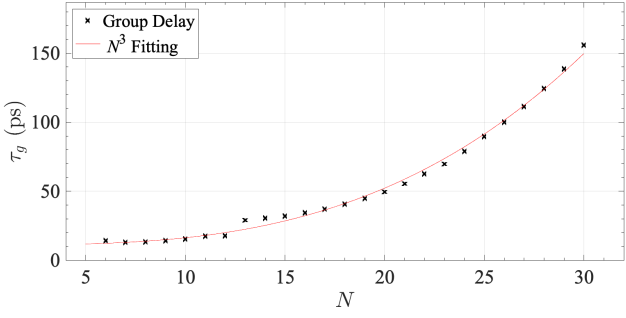

Fig. 10: Group delay of the waveguide at the resonance frequency using analytic methods for an increasing number of unit cells $N$. The fitting curve is described by $aN^3 + b$ where $a = 5.1$ fs and $b = 11.1$ ps.

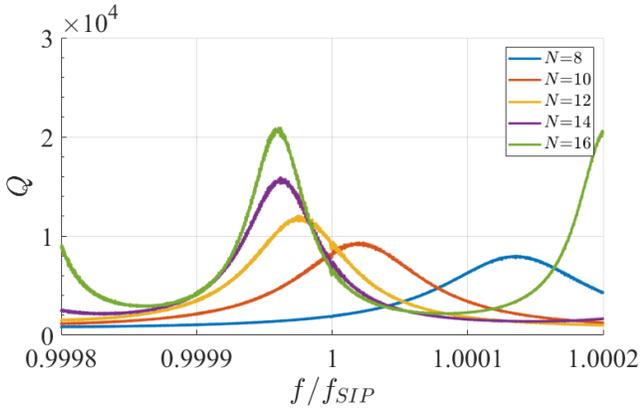

Fig. 11: Quality factor evaluated near the SIP frequency, calculated using analytic methods for waveguides with five different numbers of unit cells $N$.

## Appendix A
### Changing System Matrix Definitions

There are two common representations of the state vector in $n$-port optical waveguides: grouping the electric fields in each waveguide together as in this paper and given in Eq. (1), or grouping the field propagating direction together as in [8] and given as

$$\boldsymbol{\Psi}'(z) = \begin{pmatrix} \mathbf{E}^+(z) \\ \mathbf{E}^-(z) \end{pmatrix} = \big[\, E_1^+(z),\ E_2^+(z),\ E_3^+(z), \\ E_1^-(z),\ E_2^-(z),\ E_3^-(z)\,\big]^T \quad (16)$$

for a six-port waveguide like in this paper.

The choice of state vector determining the unit cell transfer matrix is generally arbitrary and has no effect on the eigenvalues of a matrix and thus no effect on the dispersion diagram [44]. However, depending on modeling convenience and the proposed geometry, there may be a state vector definition more beneficial when working with a given geometry. Grouping by propagation direction ($\boldsymbol{\Psi}'$) is the same notation given in [45] for their summary on physical constraints imposed on the $T$ matrix. Describing a linear change of basis transformation between the two mentioned state vector definitions proves useful in verifying reciprocal and lossless properties of the transfer matrix.

We start with this paper's transfer matrix $\underline{\mathbf{T}}$ that uses the state vector in Eq. (1). To rewrite the transfer matrix that uses the state vector in Eq. (16), we apply

$$\underline{\mathbf{T}}' = \underline{\mathbf{A}}\,\underline{\mathbf{T}}\,\underline{\mathbf{A}}^{-1} \quad (17)$$

where $\underline{\mathbf{A}}$ is the permutation matrix defined as

$$\underline{\mathbf{A}}_6 = \begin{pmatrix} 1 & 0 & 0 & 0 & 0 & 0 \\ 0 & 0 & 1 & 0 & 0 & 0 \\ 0 & 0 & 0 & 0 & 1 & 0 \\ 0 & 1 & 0 & 0 & 0 & 0 \\ 0 & 0 & 0 & 1 & 0 & 0 \\ 0 & 0 & 0 & 0 & 0 & 1 \end{pmatrix}, \quad \underline{\mathbf{A}}_4 = \begin{pmatrix} 1 & 0 & 0 & 0 \\ 0 & 0 & 1 & 0 \\ 0 & 1 & 0 & 0 \\ 0 & 0 & 0 & 1 \end{pmatrix} \quad (18)$$

for a six state vector transformation and a four state vector transformation, respectively. When changing the state vector definition in the reverse direction, we apply

$$\underline{\mathbf{T}} = \underline{\mathbf{A}}^{-1}\,\underline{\mathbf{T}}'\,\underline{\mathbf{A}}. \quad (19)$$

The $T'$ matrix, or transform of the $T$ matrix, obeys the fundamental $J$-unitary property when describing a lossless system. This property means $\underline{\mathbf{T}}^{-1}(z_2, z_1) = \underline{\mathbf{J}}\,\underline{\mathbf{T}}^\dagger(z_2, z_1)\,\underline{\mathbf{J}}^{-1}$ with $\underline{\mathbf{J}}$ defined in Eq. (20). This is similar to general stratified media in [7]. The unit cell transfer matrix must use the state vector in Eq. (16) and not in Eq. (1) (and thus use $\underline{\mathbf{T}}'$ instead of $\underline{\mathbf{T}}$) to use the definition of

$$\underline{\mathbf{J}} = \begin{pmatrix} \underline{\mathbf{1}} & \underline{\mathbf{0}} \\ \underline{\mathbf{0}} & -\underline{\mathbf{1}} \end{pmatrix} \quad (20)$$

and the conditions described in [45].

## Appendix B
### Transformation From $S$ Matrix to $T$ Matrix

We show here the transformation steps from the scattering $S$ matrix [46] to the transfer $T$ matrix. This transfer matrix uses the state vector as in Eq. (1). However, the below transformation from $S$ to $T$ only works with a two-port network if using the state vector definition in Eq. (1). To use the transformation in a more general $n$-port network, the transfer matrix must use the state vector in Eq. (16). For this paper's structure, we first determine the $S$ matrix as needed and transform it to the $T'$ matrix describing the evolution of the state vector in Eq. (16) using Eq. (22). Since we define our state vector in Eq. (1), we use Eq. (19) to determine the transfer matrix $T$.

To calculate an $n$-port transformation, the $S$ matrix and $T$ matrix are divided into four subblock matrices. For a four-port waveguide the subblocks are each of size $2 \times 2$. For the entire structure, or a six-port waveguide, the subblocks are each of size $3 \times 3$. The general representations of the scattering and transfer matrices are

$$\underline{\underline{S}}(z_1, z_2) = \begin{pmatrix} \underline{\underline{S}}_{11} & \underline{\underline{S}}_{12} \\ \underline{\underline{S}}_{21} & \underline{\underline{S}}_{22} \end{pmatrix}, \quad \underline{\underline{T}}'(z_1, z_2) = \begin{pmatrix} \underline{\underline{T}}'_{11} & \underline{\underline{T}}'_{22} \\ \underline{\underline{T}}'_{21} & \underline{\underline{T}}'_{22} \end{pmatrix}. \quad (21)$$

We now transform the $S$ matrix subblocks into the $T$ matrix subblocks through the following expressions [45]:

$$\begin{aligned} \underline{\underline{T}}'_{11} &= \underline{\underline{S}}_{21} - \underline{\underline{S}}_{22}\underline{\underline{S}}^{-1}_{12}\underline{\underline{S}}_{11}, \\ \underline{\underline{T}}'_{12} &= \underline{\underline{S}}_{22}\underline{\underline{S}}^{-1}_{12}, \\ \underline{\underline{T}}'_{21} &= -\underline{\underline{S}}^{-1}_{12}\underline{\underline{S}}_{11}, \\ \underline{\underline{T}}'_{22} &= \underline{\underline{S}}^{-1}_{12}. \end{aligned} \quad (22)$$

The reverse transformations are

$$\begin{aligned} \underline{\underline{S}}_{11} &= -\underline{\underline{T}}^{-1}_{22}\underline{\underline{T}}_{21}, \\ \underline{\underline{S}}_{12} &= \underline{\underline{T}}^{-1}_{22}, \\ \underline{\underline{S}}_{21} &= \underline{\underline{T}}_{11} - \underline{\underline{T}}_{12}\underline{\underline{T}}^{-1}_{22}\underline{\underline{T}}_{21}, \\ \underline{\underline{S}}_{22} &= \underline{\underline{T}}_{12}\underline{\underline{T}}^{-1}_{22}, \end{aligned} \quad (23)$$

where the superscript prime on the $T$' subblocks are dropped for better readability.

## Appendix C
### Chirped DBR Design and Optimization

To realize the SIP, we require field reflection along the three-way OWG to facilitate $+z$ and $-z$ propagating modes mixing. Note that in a three-way OWG with only phase delay and coupling blocks, we did not see any SIP. We achieve reflection using two distributed Bragg reflectors (DBRs) in each unit cell of the OWG. A DBR is realized by periodically changing the width of a straight waveguide for an effective modulation of the modal refractive indices. This section will discuss the DBR design and the final chirped DBR used to find the SIP.

We initially use a 20 cell DBR with a cell length of 350 nm, duty cycle between unperturbed and perturbed widths of 0.5, and a center (perturbed) width of $d = 350$ nm. A full-wave frequency domain FEM simulation gives the magnitude of $|S_{11}| = 0.567$ and $|S_{12}| = 0.815$ at 193 THz for the DBR. As customary, to remove affects from higher-order evanescent modes due to the width discontinuity, the simulation port monitor was moved an appropriate distance away from the discontinuity and then used the de-embedding function to reconstruct the phase information at the precise ends of the DBR. The variation between $S_{11}$ and $S_{22}$, and $S_{12}$ and $S_{21}$ is on the order of simulation error thresholds; indeed, the DBR is symmetric and reciprocal as assumed in Sec. II-B.

It is also useful to define a loss equation when describing the DBR. For a lossless structure, $|S_{11}|^2 + |S_{21}|^2 = 1$. This is identical to the unitary property of $S$ discussed in Sec. II-B. Therefore, we will consider losses represented by

$$L_{dB} = -10 \log(|S_{11}|^2 + |S_{21}|^2), \quad (24)$$

where, again, $S_{22}$ and $S_{12}$ are interchangeable for $S_{11}$ and $S_{21}$, respectively, and the lower the value of $L_{dB}$, the less lossy the DBR.

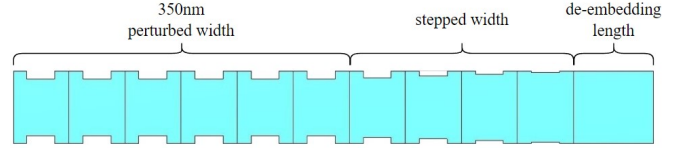

Fig. 12: Half of the mirrored chirped DBR made of 10 cells, with cross section dimensions for the six unperturbed cells given as $h = 220$ nm and $w = 450$ nm. Four cells are used for the chirped section, on each side of the chirped DBR.

This 20 cell DBR has $L_{dB} = 0.067$ dB which corresponds to approximately 1.4% power loss. All losses are due to radiation.

A solution to minimize radiation losses is to step the width difference gradually until the desired width is achieved. The radiation loss due to the initial step from 450 nm to 430 nm is significantly less than the radiation loss when you step directly from a 450 nm to 350 nm width. We optimize around a 20 nm chirping of the widths down to 350 nm. The chirped DBR (cDBR) is made by four cells on each side of the cDBR with minimum width of $\{430, 400, 375, 360\}$ nm between the unperturbed straight waveguide and fully perturbed part of the cells in cDBR. This chirping of the widths significantly reduces losses in the DBR. Half of the symmetric cDBR is shown in Fig. 12. The cDBR model has 12 periods of width 350 nm in the center. The resulting $S$ parameters of the whole cDBR (still 20 cells) are $|S_{11}| = 0.518$ and $|S_{12}| = 0.856$ at 193 THz, and $L_{dB} = 0.0029$ dB which is significantly less than the one without chirped cells. Fig. 13 shows the difference in radiation loss between the two DBR designs. Clearly the cDBR is a significant improvement over the DBR when comparing losses.

There is a tradeoff between length, perturbed width variation, and reflection coefficient for the DBR and cDBR. Using a varying width, we reduce the dB loss in the chirped DBR by over an order of magnitude. When using the chirped DBR, we also see a good agreement in the dispersion diagram calculated via the analytic model (lossless, by construction) and the full-wave model (accounting for radiation losses) in Fig. 4b. From our previous studies with arbitrary reflection coefficients (not based on full-wave simulations), we see that we can find the SIP over a wide range of reflection coefficients. Therefore, there is a considerable variety in the DBR design space depending on fabrication or size requirements.

We also note that once the cDBR was designed, we used the full-wave simulation generated $S$-parameters "as-is". The scattering matrix is normalized to the impedance of the rectangular straight OWG as in Fig. 2b. The geometric parameters of the unit cell of the three-path OWG are then optimized around the fixed cDBR to find the SIP. This was primarily done because of the complexity of fine-tuning the scattering parameters in high-accuracy simulations taking upwards of one day to complete. Changing and optimizing the other parameters (not the cDBR) in the analytic model allowed for expedited tuning and optimization for SIP behavior.

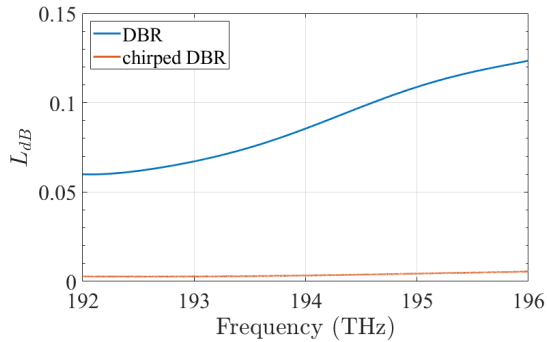

Fig. 13: dB loss calculations from Eq. (24) calculated via full-wave simulations of the scattering parameters for the DBR and chirped DBR, both made of 20 cells. As expected, the chirped DBR shows less radiation losses.

## Appendix D
## Even and Odd Mode Coupler

As stated in Sec. II-C, the unit cell transfer matrix for both the analytic and full-wave models uses a subblock derived from the even and odd mode electric fields in coupled waveguides. This appendix describes the conversion from even and odd modes to $+z$ and $-z$ propagating and gives a graphic representation of the coupling model.

We start by writing the even and odd field modes in terms of the individual electric field modes in each waveguide as

$$\mathbf{E}_e(z) = \mathbf{E}_1(z) + \mathbf{E}_2(z)$$
$$\mathbf{E}_o(z) = \mathbf{E}_1(z) - \mathbf{E}_2(z) \qquad (25)$$

where the two-dimensional wave amplitude vectors $\mathbf{E}_1$ and $\mathbf{E}_2$ are as defined before in Sec. II. It is critical to note that the even and odd modes are between both straight waveguides, with the even mode showing a strong field in the gap between the two straight waveguides and the odd mode a strong field in each straight waveguide and not in the gap between them. Similar to Eq. (4), we write the phase propagation for even and odd modes as $\Omega_e = e^{-jn_e k_0 L_c}$ and $\Omega_o = e^{-jn_o k_0 L_c}$. Here, $n_e$ and $n_o$ are the modal refractive indices for the even and odd modes, respectively, in the coupler. Then, we write the scattering matrix for even and odd modes as

$$
\begin{pmatrix} E_e^-(z) \\ E_o^-(z) \\ E_e^+(z+L_c) \\ E_o^+(z+L_c) \end{pmatrix} = \underline{\mathbf{S}}_{eo} \begin{pmatrix} E_e^+(z) \\ E_o^+(z) \\ E_e^-(z+L_c) \\ E_o^-(z+L_c) \end{pmatrix},
$$

$$
\underline{\mathbf{S}}_{eo} = \begin{pmatrix} 0 & 0 & \Omega_e & 0 \\ 0 & 0 & 0 & \Omega_o \\ \Omega_e & 0 & 0 & 0 \\ 0 & \Omega_o & 0 & 0 \end{pmatrix}. \quad (26)
$$

The modes propagate through the given length $L_c$ and result in a coupling coefficient. By substituting the two relationships in Eq. (25) into Eq. (26) and solving for $\mathbf{E}_1$ and $\mathbf{E}_2$, we determine the transfer matrix for the distributed coupler as Eq. (7) given in Sec. II-C.

We use the port analysis of the frequency domain solver in CST Studio Suite (as done with $n_w$ in Sec. II-A) to calculate the even and odd mode refractive indices. The waveguide dimensions are as in Fig. 2b with a 150 nm gap between waveguides. At 193 THz, the even and odd mode refractive indices are $n_e = 2.385$ and $n_o = 2.335$, respectively.

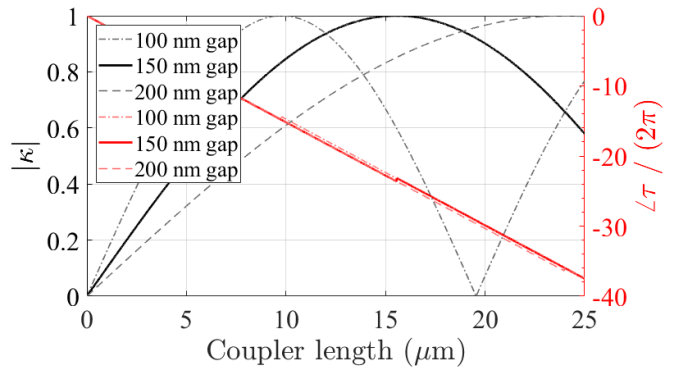

Fig. 14: Magnitude of the field coupling coefficient $|\kappa|$ and phase of the field transmission coefficient varying length $L_c$ of a distributed coupler, for three gap sizes at 193 THz.

In Fig. 14, we show the field coupling coefficient magnitude and the forward mode field transmission coefficient phase for a given length and gap size. In approximately 15 μm we see unity coupling for a 150 nm gap size. As expected, a smaller gap provides unity coupling in a shorter length and a larger gap has unity coupling in a longer length. We also see a linearly changing phase delay corresponding to the changing length.

## Acknowledgment

This material is based upon work supported by the Air Force Office of Scientific Research award numbers LRIR 21RYCOR019 and FA9550-18-1-0355.